\documentclass[aps,twocolumn,groupedaddress,preprintnumbers,nofootinbib]{revtex4-1}
\usepackage{amssymb, amsmath, siunitx}
\usepackage{hyperref}
\usepackage{slashed}
\usepackage{bbold}
\usepackage{graphicx}

\begin{document}



\vspace*{1mm}

\title{Starobinsky-Like Inflation in Dilaton-Brane Cosmology}

\author{John~Ellis$^{a}$}
\email{John.Ellis@cern.ch}
\author{Nick~E.~Mavromatos$^{a}$}
\email{nikolaos.mavromatos@kcl.ac.uk}
\author{Dimitri~V.~Nanopoulos$^{b}$}
\email{dimitri@physics.tamu.edu}

\vspace{0.1cm}
\affiliation{
${}^a$ Theoretical Particle Physics and Cosmology Group, Department of
  Physics, King's~College~London, London WC2R 2LS, United Kingdom;\\
Theory Division, CERN, CH-1211 Geneva 23,
  Switzerland
 }
 \affiliation{
${}^b$ 
 George P. and Cynthia W. Mitchell Institute for Fundamental Physics and Astronomy,
Texas A\&M University, College Station, TX 77843, USA;\\
Astroparticle Physics Group, Houston Advanced Research Center (HARC), Mitchell Campus, Woodlands, TX 77381, USA;\\
Academy of Athens, Division of Natural Sciences,
28 Panepistimiou Avenue, Athens 10679, Greece,}
 
\begin{abstract} 
We discuss how Starobinsky-like inflation may emerge from dilaton dynamics in brane cosmology scenarios based on
string theory, in which our universe is represented as a three-brane. The
effective potential may acquire a constant term from a density of effectively point-like non-pertubative
defects on the brane. Higher-genus corrections generate corrections to the effective potential
that are exponentially damped at large field values, as in the Starobinsky model, but at a faster
rate, leading to a smaller prediction for the tensor-to scalar perturbation ratio $r$.
This may be compensated partially by logarithmic deformations on the world-sheet due to
recoil of the defects due to scattering by string matter on the brane, which tend to enhance
the tensor-to-scalar ratio.  Quantum fluctuations of the ensemble of D-brane defects during the inflationary period
may also enhance the tensor-to-scalar ratio. 

\end{abstract}

%

\maketitle

The remarkable consistency of the inflationary model of Starobinsky~\cite{staro}
with the first installment of data from the Planck satellite~\cite{Planck} has triggered
great interest in the current literature, revisiting the model from various points of view,
such as no-scale supergravity~\cite{olive1,olive2}, related models including superconformal supergravity~\cite{sugrainfl},
dynamically-broken supergravity~\cite{ahm} and induced gravity~\cite{Giudice}.
The Starobinsky model obtains a de Sitter (inflationary) cosmological solution to the gravitational equations by postulating an
action in four space-time dimensions that includes a term quadratic in the scalar curvature~\cite{staro}
\begin{eqnarray}\label{staroaction}
{\mathcal S} = \frac{1}{2 \, \kappa^2 } \, \int d^4 x \sqrt{-g}\,  \left(R  + \beta  \, R^2 \right) ~,~ 
\beta = \frac{8\, \pi}{3\, {\mathcal M}^2 }~,
\end{eqnarray}
where $\kappa^2=8\pi G$, and ${\rm G}=1/m_P^2$ is Newton's (gravitational) constant in four space-time dimensions,
$m_P$ is the Planck mass, and ${\mathcal M}$ is a constant of mass dimension one, characteristic of the model. 
The scale of inflation in this model is linked to the magnitude of $\beta$ or, equivalently, to that of ${\mathcal M}$,
and the data require $\beta \gg 1$ (${\mathcal M} \ll m_P$). From a microscopic point of view, such a large
value of $\beta$ might appear somewhat surprising, and a challenge to anchor in a more complete quantum
theory of gravity~\footnote{We note, however, that quantum-gravity corrections to (\ref{staroaction}) have been considered recently
in~\cite{copeland}, from the point of view of an exact renormalisation-group (RG) analysis~\cite{litim}, 
with the conclusion that the largeness of the $R^2$ coupling, required for agreement with
inflationary observables~\cite{Planck}, is naturally ensured by the presence of an asymptotically-free ultra-violet fixed
point.}.

\begin{figure}[h!!!]
\centering
		\includegraphics[width=0.45\textwidth]{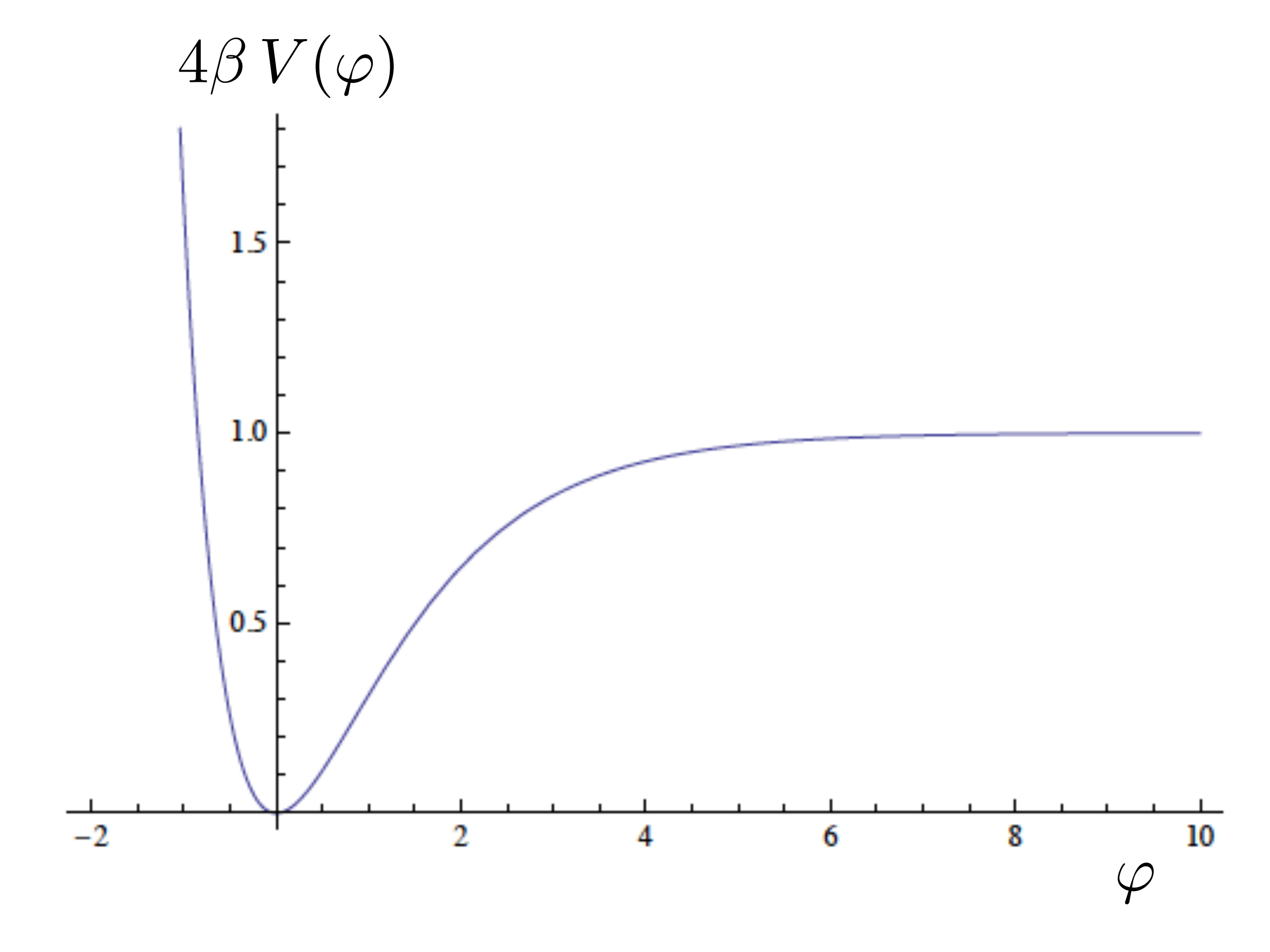}
		\caption{The effective  potential (\ref{staropotent}) of the collective scalar field $\varphi$ in the Starobinsky model for inflation (\ref{staroaction}).}
\label{fig:potstar}	
\end{figure}

Although the Starobinsky model might appear not to contain any fundamental scalar field
that could be the inflaton, it is in fact conformally equivalent to ordinary Einstein gravity coupled to a scalar field
with an effective potential that drives inflation~\cite{whitt}. To see this, one first linearizes the $R^2$ terms in 
(\ref{staroaction}) by means of an auxiliary Lagrange-multiplier field $\tilde \alpha (x)$,
then rescales the metric by a conformal transformation and redefines the conformal scalar field 
so that the theory is written in terms of canonically-normalized Einstein and scalar-field terms:
\begin{eqnarray}\label{confmetric}
g_{\mu\nu} \rightarrow g^E_{\mu\nu} & = & \left(1 + 2 \, \beta \, {\tilde \alpha (x)} \right) \, g_{\mu\nu} ~, \\
 \tilde \alpha \left(x\right) \to \varphi (x) & \equiv & \sqrt{\frac{3}{2}} \, {\rm ln} \, \left(1 + 2\, \beta \, {\tilde \alpha \left(x\right)} \right)~,
\end{eqnarray}
so that
\begin{align}\label{steps}
	&\frac{1}{2\kappa^2} \, \int d^4 x \sqrt{-g}\,  \left( R  + \beta  \, R^2 \right)  \nonumber \\
  	&\hookrightarrow  \frac{1}{2\kappa^2}\, \int d^4 x \sqrt{-g}\,  \left(  \left(1 + 2\, \beta \, \tilde \alpha \left(x\right) \right) \, R  -  \beta  \, {\tilde \alpha (x)}^2 \right)\nonumber  \\
    & \hookrightarrow \frac{1}{2\kappa^2}\,\int d^4 x \sqrt{-g^E}\,  \left(R^E +  g^{E\, \mu\, \nu} \, \partial_\mu \, \varphi \, \partial_\nu \, \varphi - V\right(\varphi\left) \right)~,\nonumber \\
\end{align}
with the effective potential $V(\varphi)$ given by:
\begin{eqnarray}\label{staropotent}
 V(\varphi ) = \frac{\left( 1 - e^{-\sqrt{\frac{2}{3}} \, \varphi } \right)^2}{4\, \beta} \, 
  = \frac{3 {\mathcal M}^2 \, \Big( 1 - e^{-\sqrt{\frac{2}{3}} \, \varphi } \Big)^2}{32\, \pi }  \,,
\end{eqnarray}
which is plotted in Fig.~\ref{fig:potstar}. Here we see explicitly that $\beta \gg 1$ or, equivalently,
$\mathcal{M} \ll m_P$ corresponds to a potential with magnitude $V \ll m_P^4$, a necessary
condition for successful inflation. This formulation of the Starobinsky model in terms of
a scalar field lends itself more naturally to inclusion in a broader theoretical context, which
has been the thrust of much of the recent research~\cite{olive1,olive2,sugrainfl,ahm,Giudice}.

At large values of the dimensionless scalar field $\varphi$, i.e.,
values of the dimensionful field $\phi = \kappa^{-1}\, \varphi$ that are large compared to the Planck scale, 
the potential (\ref{staropotent}) is sufficiently flat to produce phenomenologically acceptable inflation, 
with the scalar field $\varphi$ playing the role of the 
inflaton. Thus, the region of the positive-definite Starobinsky potential (\ref{staropotent}) that is relevant for inflation
is that where the constant and the $e^{-\sqrt{2/3} \varphi}$ terms are dominant.

With this in mind, one is led to consider phenomenological generalizations of (\ref{staropotent}) of the form~\cite{olive2}:
\begin{equation}\label{phenostar}
\tilde V \equiv \frac{1}{2\kappa^2}\, V = \frac{A}{2 \kappa^2} \Big( 1 - \delta\, e^{-B \varphi} + \dots \Big) \, ,
\end{equation}
where $1 \gg A > 0$, and $\delta $ and $B$ may be treated as free parameters that are allowed to vary from the original
values $\delta =2$, $B=\sqrt{2/3}$ in our normalization for the Einstein-frame Starobinsky
action, cf, (\ref{steps}), and the dots represent possible higher-order terms that are $\mathcal{O}(e^{-2 B \varphi})$. 
Using the following standard expressions for inflationary observables in the slow-roll approximation
\begin{eqnarray}\label{slowrollparam}
\epsilon \; = \; \frac{1}{2} M^2_{\rm Pl} \Big(\frac{V^{\prime}}{V}\Big)^2 & , &  \eta \; = \; M^2_{\rm Pl} \Big(\frac{V^{\prime\prime}}{V}\Big)~, \nonumber \\
n_s \; = \; 1 - 6\epsilon + 2\eta & , & r \; = \; 16 \epsilon~, \, \, \nonumber \\
N_\star & = & - M_{\rm Pl}^{-2} \, \int_{\varphi_i}^{\varphi_e} \frac{V}{V^\prime} d\varphi ~,
\end{eqnarray}
where $M_{\rm Pl} = M_{\rm P}/\sqrt{8\pi}$ is the reduced Planck mass 
and $\varphi_{i(e)}$ indicate the values of the inflaton at the beginning (end) of the inflationary era,
at leading order in the small quantity $e^{-B\varphi}$~\cite{olive2}:
\begin{eqnarray}
&& n_s = 1 - 2 B^2 \, \delta \,  e^{-B \varphi}~, \, r = 8 B^2 \delta^2 \, e^{-2 \, B \varphi}, \, N_\star = \frac{1}{B^2\, \delta} e^{B \varphi} \, , \nonumber \\
&& {\rm yielding} \quad \, n_s = 1 - \frac{2}{N_\star} , \quad r = \frac{8}{B^2\, N_\star^2}~,
\label{crucial}
\end{eqnarray}
where $N_\star $ is the number of e-foldings during the inflationary phase. Requiring 
$N_\star = 54 \pm 6$ yields the characteristic predictions $n_s = 0.964 \pm 0.004$, and the Starobinsky-model choice 
$B=\sqrt{2/3}$ yields $r= 0.0041^{+0.0011}_{-0.0008}$, which is highly
consistent with the Planck data~\cite{Planck}. The question then arises how one could deviate from the characteristic 
Starobinsky predictions. 

\medskip
In this note we answer this question within the context of inflationary scenarios derived from brane cosmology and
non-critical string. As we shall show, this approach predicts a different value of the coefficient $B = \sqrt{2}$, 
a possibility suggested previously in the framework of no-scale supergravity~\cite{olive2}, but due here to the structure
of the string genus expansion. Our approach
also suggests the possibility that the coefficient $\delta$ in (\ref{phenostar}) may depend linearly on $\varphi$
due to the appearance of a logarithmic operator.
The prediction for $B$ suggests a {\it smaller} value of $r$ than in the Starobinsky model, which could be
partially compensated by an enhancement due to the possible linear dependence of $\delta$ on $\varphi$.

\medskip
We discussed in~\cite{string} a scenario for inflation within non-critical string theory
that involved colliding brane universes at early epochs. The brane collision caused the string excitations on the 
observable Universe brane to be described by super-critical Liouville string, 
with {\it positive} central charge deficit in the standard terminology~\cite{aben}. In that case, 
a de Sitter inflationary phase was obtained due to an anti-alignment in field space of  an 
antisymmetric tensor field equivalent in four dimensions to a massless axion-like field
scaling linear with the Robertson-Walker (RW) Einstein-frame time~\cite{aben} with a dilaton field
scaling logarithmically with the RW time. The presence of the brane was essential for ensuring the 
appropriate four-dimensional structures that lead to inflation in this scenario. This brany Liouville 
model of inflation is also consistent with the Planck data~\cite{string}, but has more free parameters 
than the Starobinsky model, since it involves two fields: one scalar (dilaton) and one pseudoscalar (axion-like field).
This multifield inflationary scenario can be studied in the same way as the complex Wess-Zumino inflationary
model~\cite{WZ}.

This particular {\it super}-critical string model did not yield an effective potential of the type (\ref{phenostar}),
but we argue here that  potentials of the form (\ref{phenostar}) with 
the non-Starobinsky value $B = \sqrt{2}$ (in our normalization (\ref{steps})
occur generically in dilaton/{\it sub}-critical string cosmologies, i.e., scenarios
with {\it negative} central charge deficits in the terminology of \cite{aben}, in the presence of branes. 
The branes provide a cosmological constant $A$ that is non-perturbative in the string genus expansion,
and independent of the Starobinsky inflaton field $\varphi$, 
which is identified (up to a minus sign) with the canonically-normalized) Liouville/dilaton of 
the closed-string multiplet in the Einstein frame, i.e., $\varphi = -\Phi$, where $\Phi$ is the dilaton
and the string coupling is $g_s = e^\Phi = e^{-\varphi}$. 

In this {\it sub}-critical string approach, the generalized Starobinsky potential 
(\ref{phenostar}) arises from non-perturbative string theory contributions involving brane worlds of the form
\begin{equation}\label{nonpert}
\tilde V = A\, \times \, \mathcal{O}(e^{-F_D/g_s}) \, ,
\end{equation}
where the coefficient $A$ is independent of the string coupling $g_s$, whereas the
constant $F_D > 0$ is due to 
contributions from D-branes localized at some point in the (higher-dimensional) space-time of string theory. 
In addition to $c=1$ subcritical strings~\cite{kazakov},
contributions of the form (\ref{nonpert}) are known to appear in
models with central charge $c=26$ that feature a cigar-type two-dimensional black-hole metric~\cite{witten} 
with D-branes localized at the tip of the cigar, corresponding to $\frac{SL(2)_k}{U(1)} \times \frac{SU(2)}{U(1)}$
backgrounds. The coefficients $A$ and $F_D$ in such models have been explicitly computed using matrix models, 
which are used to represent a non-perturbative version of $c=1$ strings.
The important issue for our purpose is how to identify contributions to the effective action on the brane
that take the form of a cosmological constant $A$ term that is independent of the dilaton
in the Einstein frame, which we now discuss. 

We first consider the case of a higher-dimensional closed string propagating in a background
containing graviton and dilaton fields $G_{\mu\nu}, \Phi$.
As is well known, to lowest order in the \emph{perturbative} world-sheet genus $h \equiv 2-\chi$, 
corresponding to the first term in an expansion in powers $g_s^{-2+\chi}=(e^\Phi)^{-2 + \chi}$
of the closed-string coupling, namely the world-sheet sphere with $\chi=0$,
the effective target-space action in the $\sigma$-model or string-frame is given by:
\begin{eqnarray}\label{eas}
 S_{string} &=&  \frac{1}{2\kappa^2} \int d^Dx \sqrt{-\tilde g} e^{-2\Phi}\Big ( -\frac{2(D-26)}{3\alpha^\prime} \nonumber \\
 &+& R(\tilde g) + 4 (\partial_\mu \Phi)^2 + {\mathcal O}(\alpha^\prime ) \Big) ~,
\end{eqnarray}
where $\kappa$ is the higher-dimensional gravitational coupling constant, expressed in terms of 
the string scale $M_s = (\alpha^\prime)^{-1/2}$ via
$\kappa^2 = 8\pi (\sqrt{\alpha^\prime})^{D-2} {\mathcal V}^{-1}= 8\pi M_s^{-(D-2)} {\mathcal V}^{-1}$ where  
${\mathcal V} = \Pi_{i}^n (R_i/\sqrt{\alpha^\prime})$ is the (dimensionless) compactification volume factor,
and $R_i, i=1, \dots n = 26-D$ are the compactification radii where
$D$ is the number of the large (uncompactified) dimensions of the string. 
Transforming to the Einstein frame, given by
\begin{equation}\label{einst}
g_{\mu\nu} = e^{-\frac{2\Phi}{D-2}} \, \tilde g_{\mu\nu}~,
\end{equation}
the effective action (\ref{eas}) becomes
\begin{eqnarray}\label{eae}
 S_{string} &=&  \frac{1}{2\kappa^2} \, \int d^Dx \sqrt{-\tilde g} \Big ( - e^{\frac{4\Phi}{D-2}}\, \frac{2(D-26)}{3\alpha^\prime} \nonumber \\
 &+& R(g) - \frac{4}{D-2}\,  (\partial_\mu \Phi)^2 + {\mathcal O}(\alpha^\prime ) \Big) ~,
\end{eqnarray}
on the world-sheet sphere. In the case of superstrings, one must replace $2(D - 26)/3 \alpha^\prime$ in
equations (\ref{eas}, \ref{eae}) by $(D - 10)/\alpha^\prime$.

We now consider non-perturbative effects, specifically punctures in the $D$-dimensional space-time due to $D$-brane defects. 
The simplest scenario is to consider effective D0-branes that, depending on the string theory considered, 
may be either truly point-like (as in Type-IIA theory) or effectively point-like, e.g., D3-branes
compactified on 3-tori, which, from the point of view of a low-energy three-space observe,
would be effectively point-like defects~\cite{westmuckett,dfoam}. In the presence of D3-brane universes
moving in the bulk space, the cosmology of such populations of D0-brane defects and
their interactions with ordinary string matter on the brane (recoil) have been considered in~\cite{dfoam}.
For our purposes here we consider the simplest scenario in which the closed-string
backgrounds are non-trivial on the D3 brane but are trivial in the bulk, so we set $D=4$ in the effective action 
(\ref{eae}). The rest of the space-time dimensions are assumed either to have been compactified appropriately, 
or to represent bulk dimensions in which the space-time is flat.
This bulk space-time is assumed to be punctured by (effectively) point-like D0-brane defects
as mentioned previously, and we consider the inflationary era of such a string Universe. 

The D0-branes have mass $M_s/g_s$ where $g_s = {\overline g}_s \, e^\Phi (t, \vec x)$ is the (fluctuating) string scale. 
If they have a density per unit three-volume $n$ on the brane universe, 
they can contribute to the vacuum energy of the brane the following non-perturbative term
in the $\sigma$-model frame:
\begin{equation}
\Lambda_{\rm D-branes} = - \int d^4 x \sqrt{-\tilde g} e^{-\Phi} M_s n ~.
\end{equation}
In the Einstein frame (\ref{einst}), this becomes 
\begin{equation}\label{einstd0}
\Lambda^E_{\rm D-branes} = - \int d^4 x \sqrt{-g} M_s \tilde n ~,
\end{equation}
where we have taken into account the scaling of the density $ \tilde n = e^{-3\Phi} n $ with the proper three-volume.

The inflationary phase is one in which there is a constant density of D0-branes 
(and thus brane creation) in $D=4$, which may arise from an influx of D-particles entering the D3 brane from the bulk,
that compensates the density dilution due to the exponential expansion of the brane universe space-time.
In this case the D0-contribution to the brane vacuum energy is constant and dilaton-independent 
in the Einstein frame on the brane.

In the following, we perform the redefinition 
\begin{equation}\label{redef}
\Phi \to -\Phi \, ,
\end{equation}
which corresponds to the well-known strong-weak coupling string duality $g_s \to g_s^{-1}$.
Combining the terms (\ref{eae}) and (\ref{einstd0}), making this duality
transformation, and normalizing the dilaton kinetic term as in (\ref{steps}), i.e., 
redefining $\varphi = \sqrt{2}\Phi$, we arrive at the following effective target-space action 
on the brane world:
\begin{eqnarray}\label{final}
S_{\rm eff} &=& \frac{1}{2\kappa^2} \int d^4 x \sqrt{-g} \Big(R - (\partial_\mu \varphi)^2 \nonumber \\
&-& 16 \pi\, \frac{\tilde n}{{\overline g}_s\, M_s \mathcal V} + \frac{44}{3 \alpha^\prime} e^{-\sqrt{2}\varphi} 
 + \dots  \Big) \, ,
\end{eqnarray}
where the reader is reminded that $\kappa^2 = 8\pi \alpha^\prime {\mathcal V}^{-1} = 
8\pi M_s^{-2} {\mathcal V}^{-1} \equiv M_{\rm Pl}^{-2} $ is the four-dimensional reduced Planck mass, and the dots 
denote higher-order terms coming, e.g., from higher-order terms in the string genus
expansion~\footnote{For example, massless vector fields on the 3-brane Universe
with Maxwell field strengths and a Born-Infeld world-volume action could make contributions
to the vacuum energy $\propto e^{-3\varphi/\sqrt{2}}$ in the Einstein frame.}. We see explicitly
that the effective potential (\ref{final}) has a form similar to that of the Starobinsky model (\ref{phenostar}) 
at large $\varphi$, with the correspondence: 
\begin{equation}
A \; = \; 16 \pi \, \frac{\tilde n}{{\overline g}_s\, M_s {\mathcal V}} ~, ~A\delta \; = \; \frac{44}{3 } M_s^2 ~, ~ \, B \; = \; \sqrt{2} ~.
\label{correspondence}
\end{equation}
In the case of superstrings with critical target-space-time dimension 10, 
the relation $A \delta = 44 M_s^2/3$ in (\ref{correspondence}) is replaced by $A \delta = 6 M_s^2 $.
Inflation occurs for $\varphi \gg 1$, 
and the exit from inflation occurs small $\varphi$ around zero. Parameters of the model
such as the constant $n$ and the higher-order terms provide the flexibility to cancel the central 
charge deficit when $\varphi =0$. On the other hand, 
for negative $\varphi$ the terms of higher order in $\mathcal{O}(e^{-\varphi/\sqrt{2}})$ dominate.

Using the standard expressions (\ref{slowrollparam}) for the slow-roll parameters
we find that
\begin{equation}
n_s \; = \; 1 - \frac{2}{N_\star}\, , \; r \; = \; \frac{4}{N_\star^2} \, .
\label{nsrpred}
\end{equation}
The prediction for $n_s$ is the same as in the Starobinsky model, but the prediction
for $r$ is a factor of 3 smaller, due to the larger value of $B$ in this brane model, as also
found in a no-scale supergravity model in~\cite{olive2}.

The measured magnitude of primordial scalar density fluctuations
provides the following constraint on the effective inflationary potential:
\begin{equation}\label{constr}
\Big( \frac{\tilde V}{\epsilon} \Big)^{1/4} = 0.0275 \, M_{\rm Pl} \, .
\end{equation}
We can then use the first relation in (\ref{correspondence}), the expression 
$M_{\rm Pl}^{2} = M_s^{2} {\mathcal V}/8\pi$ and a typical value of $N_\star = 54$ to obtain an interesting
constraint on the parameters of the model, namely
\begin{equation}
\frac{\tilde n}{{\overline g}_s\, M_s^3 {\mathcal V}^2} \; \simeq \; 8 \times 10^{-14} \, .
\label{relation}
\end{equation}
In ``old'' superstring models one would expect ${\overline g}_s \sim 0.8$, 
corresponding to ${\tilde n}/{\mathcal V}^2 = {\cal O}(6 \times 10^{-14})M_s^3$. Assuming that
six dimensions are compactified with a typical scale $R$ so that ${\mathcal V} \equiv R^6$, 
the constraint (\ref{relation}) is respected for
${\tilde n} \sim {\cal O}(1) \times M_s^3$ and $R = {\cal O}(10)$. In this way, measurements of
inflationary observables may provide interesting insights into string dynamics.

An interesting generalization of (\ref{phenostar}) is natural in the formulation of $c=1$ Liouville string theory 
used above. In this theory, the world-sheet conformal algebra is 
logarithmic~\cite{kogan}, i.e., it includes a pair of operators
that are described by world-sheet deformations $\mu e^{\alpha \varphi} $ and $\mu^\prime \varphi e^\alpha \varphi$, 
where $\alpha$ is the appropriate Liouville-dressing dimension and $\varphi$ the Liouville mode. This
possibility also occurs in higher-dimensional situations, for instance in the above-mentioned
example of the propagation of strings in an ensemble of D0-branes~\cite{recoil}: the
recoil of a brane defect during its interaction with the string is described by a
logarithmic pair of operators of the above form, 
with the constants $\mu$ and $\mu^\prime$ playing the r\^oles of the position and momentum of the recoiling D0 brane. 
Thus, the incorporation of D-brane recoil in the sub-critical string model described above, 
as a result of the interaction of matter strings with the background of the ensemble of D0-branes,
modifies the potential (\ref{phenostar}) to become
\begin{eqnarray}\label{generalization}
V_{\rm recoil}^{\rm d-brane} = A - \tilde B e^{-\sqrt{2}\varphi} - C \varphi \, e^{-\sqrt{2}\varphi}~,
\end{eqnarray}
where $A,\tilde B,C$ are model-dependent constants, and we restrict ourselves to leading order in 
$e^{-\sqrt{2}\varphi}$ for large positive $\varphi$. 
This potential has a general form that is qualitatively similar to that of the original 
Starobinsky model, provided that $A,\tilde B $ and $C$ are all
positive. 

We are led to consider the inflationary predictions of a more general case,
in which the exponent is arbitrary as in (\ref{phenostar}), i.e.,
\begin{eqnarray}\label{generalization2}
V_{\rm recoil}^{\rm d-brane} = A - e^{-B\varphi} \Big( \tilde B - C \varphi \Big) + \dots \,, ,
\end{eqnarray}
where  the $\dots$ indicate higher orders in $e^{-B\varphi}$, and $B=\sqrt{2}$ in our D-foam case
({\ref{generalization}). 
Using again the slow-roll formulae (\ref{slowrollparam}), we again find that
\begin{equation}
n_s \; = \; 1 - \frac{2}{N_\star}
\label{nspred}
\end{equation}
as in the Starobinsky model., with no explicit dependence on the
parameters $A, \tilde B$ and $C$. 
On the other hand, we also find that
\begin{equation}
r \; = \; \frac{8}{B^2 N_\star^2} \left( 1 + \frac{2}{\ln N_\star} \right) \, + \dots ,
\label{rpred}
\end{equation}
where the logarithmic correction
is due to the prefactor linear in $\varphi$ in (\ref{generalization2}), and the dots represent
terms that are formally of higher order in $1/\ln N_\star$.
The leading correction in (\ref{rpred}) is numerically significant: for $N_\star = 54 \pm 6$, 
corresponding to $n_s =0.964 \pm 0.004$, one has $\ln N_\star = 3.99^{+ 0.10}_{- 0.12}$,
so the tensor-to-scalar ratio $r$ would be {\it enhanced} by $\sim 50$\% compared to the value with a constant
prefactor, as in the original Starobinsky model. However, the higher-order terms
represented in by the dots in (\ref{rpred}) are potentially important and depend, in general
on the relative magnitudes of the coefficients $\tilde B$ and $C$ in (\ref{generalization2}).

One could also consider a further generalization of the Starobinsky model, in which the 
potential has the form
\begin{eqnarray}\label{generalization3}
V_{\rm recoil}^{\rm d-brane} = A - C_n \varphi^n e^{-B\varphi} + \dots \,, ,
\end{eqnarray}
although this is not suggested by our particular stringy model, and
the dots represent possible terms of lower order in $\varphi$. It is easy to check that in this case the prediction (\ref{nspred}) is still unchanged,
whereas the prediction (\ref{rpred}) for the tensor-to-scalar ratio is modified to
\begin{equation}
r \; = \; \frac{8}{B^2 N_\star^2} \left( 1 + \frac{2 n}{\ln N_\star} + \dots \right) \, .
\label{rpredn}
\end{equation}
It is clear that a specific model for the non-leading terms in (\ref{generalization3})
would be needed to make any definite prediction, but (\ref{rpredn}) strengthens the
basic point of (\ref{rpred}), namely that the tensor-to-scalar ratio $r$ may be
enhanced significantly compared to models in which the subasymptotic corrections
are purely exponential. We note that potential terms of the form (\ref{generalization3}), but
without the constant $A$ term, that also lead to an enhanced 
scalar-to-tensor ratio $r$, have been suggested in \cite{li}, 
in the context of supergravity models with broken shift symmetries. 
The presence of the constant term $A$ in our case 
leads to different predictions for the slow-roll parameters.

In conclusion: the class of brane cosmology models discussed here offers interesting
alternatives to the Starobinsky model. The faster exponential approach to the
asymptotic constant value of the potential leads {\it a priori} to a reduced prediction
for the tensor-to-scalar perturbation ratio $r$, as seen in (\ref{nsrpred}). On the other hand, recoils of the brane
defects leading to a logarithmic correction to the effective potential (\ref{generalization2})
may compensate partially for this suppression, as seen in (\ref{rpred}). This example
reinforces the interest in improving the experimental sensitivity to $r$, which may
provide interesting insights into string and brane phenomenology. In this connection,
it would be interesting to develop further the comparisons, connections and contrasts with other recent
formulations of Starobinsky-like inflationary models~\cite{olive1,olive2,sugrainfl,ahm}. 

\begin{center}
\textbf{NOTE ADDED}
\end{center}

After this paper was submitted for publication and reviewed, 
the BICEP2 Collaboration announced strong evidence for gravitational waves at the time of the last scattering
with $r = 0.16^{+0.06}_{-0.05}$ after dust subtraction~\cite{Bicep2}.
If confirmed by subsequent experiments, such a large value of $r$ would
exclude conventional Starobinsky-type inflationary potentials.  However, in the context of our 
modified Starobinsky models (\ref{phenostar}) and (\ref{rpredn}), large values of $r$ compatible with the BICEP2 
data can be obtained if the exponent $B$ is much smaller than the conventional Starobinsky value $B_{\rm Star}=\sqrt{2/3}$. 
For instance, in the case of (\ref{phenostar}) the BICEP2 central value $r \sim 0.16$ is obtained for 
$N_\star \sim 50$ and $B \sim 0.14$.

On the other hand, our string theory considerations point towards values of $B$ that are larger than
in the conventional Starobinsky model, e.g., $B=\sqrt{4/(D-2)}=\sqrt{2}$ for the case of $D=4$ 
large uncompactified dimensions considered here, leading to a much smaller value of $r$.
Indeed, even if the dilaton lives in the maximum number $D=10$ of uncompactified space-time dimensions, 
the resulting value of $B=1/\sqrt{2}$ is much larger than required to yield $r=0.16$. 

However, we note here that quantum fluctuations of the D-particle defects
are independent sources of gravitational (tensor) perturbations.
We assumed above that the dominant r\^ole of such fluctuations was simply to provide a cosmological constant term 
in the potential (\ref{phenostar}). However, such D-particle fluctuations yield, in general
effective vector field degrees of freedom associated with stochastic fluctuations in the corresponding recoil velocities
of the D-particles as they interact with the (closed) string degrees of freedom representing bulk space-time gravitons. 
The coupled dynamics of such fluctuating defects with the space-time metric is complicated~\cite{yusaf}. 
Nevertheless, it was argued in~\cite{yusaf} that there are growing modes in such systems
that may result in the formation of large-scale structures on the D3 Brane universe at late epochs. 
It is therefore possible that, during the inflationary period, when it is assumed that there is an influx of 
D-particles from the bulk into the brane world, so as to ensure a constant density of D-particles and thus a 
cosmological constant contribution in (\ref{phenostar}), quantum fluctuations
in the flux of D-particles into the space-time may enhance the tensor-to-scalar-perturbations ratio
sufficiently to agree with the BICEP2~\cite{Bicep2}.

This scenario will be explored more fully elsewhere~\cite{emn2}, but it
would constitute a match of the  nice features of the absence of non-Gaussianities and of the running of the spectral index,
induced by the large-field flatness of the Starobinsky-like dilaton inflationary potential in the model, 
with additional quantum effects of D-particles at the end of the dilaton inflation that enhance the large value of $r$.

\section*{Acknowledgements}

The work  of J.E. and 
N.E.M. was supported in part by the London Centre for Terauniverse Studies (LCTS), using funding from the European Research Council via the Advanced Investigator Grant 267352 and by STFC (UK) under the research grant ST/J002798/1.
That of D.V.N.  was supported in part by the DOE grant DE-FG03-95-ER-40917.

  \end{document}